# Secure Routing and Data Transmission in Mobile Ad Hoc Networks

Waleed S. Alnumay and Uttam Ghosh

Computer Science Department, King Saud University, Riyadh, Saudi Arabia
Dept. of E&ECE, Indian Institute of Technology, Kharagpur-721302, India

*ABSTRACT*

*In this paper, we present an identity (ID) based protocol that secures AODV and TCP so that it can be used in dynamic and attack prone environments of mobile ad hoc networks. The proposed protocol protects AODV using Sequential Aggregate Signatures (SAS) based on RSA. It also generates a session key for each pair of source-destination nodes of a MANET for securing the end-to-end transmitted data. Here each node has an ID which is evaluated from its public key and the messages that are sent are authenticated with a signature/ MAC. The proposed scheme does not allow a node to change its ID throughout the network lifetime. Thus it makes the network secure against attacks that target AODV and TCP in MANET. We present performance analysis to validate our claim.*

*INDEX TERMS*

*MANET, AODV, TCP, Signature, Attacker, Security*

## 1. INTRODUCTION

A *mobile ad hoc network* (MANET) is a collection of two or more nodes equipped with wireless communications and networking capability. The nodes within the radio range can immediately communicate with each other. The nodes that are not within each other's radio range can communicate with the help of intermediate nodes where the packets are relayed from source to destination. Each node should be configured with an unique identity to ensure the packets correctly routed with the help of a routing protocol of a MANET.

MANETs have distinct advantages over traditional networks: (a) it can be easily set up and dismantled; (b) it is a cost-effective solution for providing communication in areas where setting up fixed infrastructures is not a suitable option constraints such as geographical location, financial implications, etc; (c) it can be set up in emergency situations (e.g., rescue mission). A node requires authentication for secure information exchange and to avoid the security threats. However, establishing secure communication in a MANET is particularly challenging task because of the following issues: (a) shared wireless medium; (b) no clear line of defense; (c) self-organizing and dynamic network; (d) most of the messages are broadcasted; (e) messages travel in a hop-by-hop manner; (f) nodes are constrained in terms of computation and battery power. In this paper, we focus on the problem of secure route discovery and data transmission in an independent MANET.

Routing protocols in a MANET can be classified into three categories based on the underlying routing information update mechanism employed: reactive (on-demand), proactive (table driven) and hybrid. In reactive routing protocols, nodes find routes only when they must send data to the destination node whose route is unknown. *Ad hoc On-demand Distance Vector* (AODV) [1] and *Dynamic Source Routing* (DSR) are under this category. On the other hand, in proactive protocols such as *Destination Sequenced Distance Vector* (DSDV), nodes periodically exchange topology information, and hence nodes can obtain route information any time they must send data. Hybrid routing protocols like *zone routing protocol* (ZRP) combine the best features of both reactive and proactive routing protocols. Each node uses proactive routing protocols to reach nodes within certain geographical area (zone), and reactive routing protocols for the rest. The reactive routing protocols are found to be more efficient in a dynamically changing topology like MANET. Under reactive routing, AODV is the most popular and is currently being researched actively. *Internet engineering task force* (IETF) has made AODV as the standard routing protocol for MANET [2]. Therefore, in this paper we have investigated and proposed improvements in AODV routing protocol.

AODV is a reactive protocol that provides route on demand basis between nodes very efficiently. It floods the route request (*RREQ*) message throughout the network at the time of route discovery process. Therefore, the RREQ message reaches the destination node and reacts with a route reply message (*RREP*). The *RREP* is sent as a unicast, using the path towards the source node established by the *RREQ*. After the successful route discovery process, data packets can be delivered from the source to the destination node and vice versa. However, it does not provide any authentication or data protection mechanism.
As a result following are the security threats [3] that are associated with AODV:

- **Attacks using modification**

  – *Redirection by modifying the route sequence number:* AODV uses monotonically increasing sequence numbers to discover and maintain the routes for a destination. A malicious node may redirect the traffic through itself by advertising with a higher destination sequence number than the actual one.
  – *Redirection by modifying the hop count:* As AODV uses the hop count field to determine a shortest path, a malicious node may divert the traffic through itself by resetting the hop count value to a smaller value.
  – *Denial-of-service by altering source/ destination:* A denial-of-service attack can be launched in AODV by modifying the source or destination address of a packet. As a result, traffic may be dropped, redirected to a different destination
  – or to a longer route to reach to destination that causes unnecessary communication delay.
  – *Tunneling:* In a tunneling attack, two or more malicious nodes may collaborate to encapsulate and exchange routing messages between them along existing data routes. As a result, the destination node falsely believes that the shortest route from the source is through these collaborating nodes and wrongly sets the path through them.

- **Impersonation attacks** In this attack, a malicious node changes its identity (such as IP address or MAC address) to a authorized node in the outgoing packets. The misbehaving node can change the topology of the network or isolate any authorized node from the network.
- **Attacks using fabrication**

  – *Falsifying route error message:* AODV implements path maintenance to recover broken paths when nodes move. If the destination node or an intermediate node along an active path moves, the node upstream of the link break sends a route error message along the reverse path toward the source node. A malicious node may send false route error message to the source node. As a result, the source node re-initiates the route discovery process by broadcasting a route request message.

Recently, a number of secure routing protocols have been proposed [2]–[7]. However, secure routing protocols alone ensure the correctness of the route discovery, cannot guarantee secure data delivery at transport layer of the protocol stack. An intelligent attacker can hide itself at the time of route discovery to place itself on a route. Later it can start dropping, forging, misrouting and injecting of data packets. Transmission Control Protocol (TCP) is one of the transport layer protocol which provides end to-end connection, reliable delivery of data packets, flow control, congestion control and end-to-end connection termination. However, it cannot provide any security mechanism and following are the attacks [8] in this layer that can be seen in MANET:

- SYN flooding attack: In SYN flooding attack, an attacker creates a large number of half-opened TCP connections with a victim node but never completes the handshake to fully open the connection. During SYN flooding, the attacker sends a large amount of SYN packets to the target node, spoofing the return address of the SYN packets. When the target machine receives the SYN packets, it sends out SYN-ACK packets to the sender and waits for ACK packet. The victim node stores all the SYN packets in a fixed-size table as it waits for the acknowledgement of the three-way handshake. These pending connection requests could overflow the buffer and may make the system unavailable for long time. Figure 1 (a) shows the normal connection establishment using three-way handshaking (when node M behaves normally) and SYN flooding attack (when node M behaves maliciously).

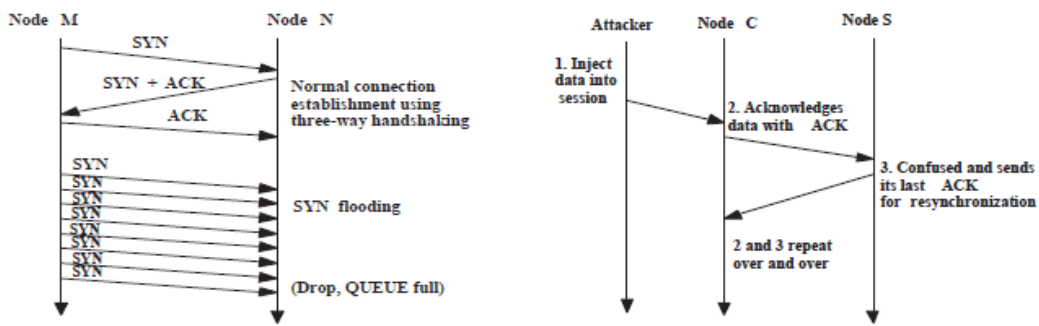

Fig. 1. (a) Normal connection establishment using three-way handshaking and SYN flooding attack; (b) ACK Storm.

- **Session Hijacking:** All the communications are authenticated only at the beginning of session setup. The attacker may take the advantage of this and commit session hijacking attack by spoofing the IP address of target machine and determining the correct sequence number. Subsequently it performs a DoS attack so that the target system becomes unavailable for a certain period of time. The attacker can now continue the session with the other system as a legitimate system.

- **ACK Storm:** The attacker launches a TCP session hijacking attack at the beginning and it then sends injected session data to node C. Node C then acknowledges the received data with an ACK packet to node S. Node S is confused as the packet contains an unexpected sequence number. Therefore, it tries to re-synchronize the TCP session with node C by sending an ACK packet that contains the intended sequence number. But the steps are followed again and again and results in TCP ACK storm which is shown in Figure 1 (b).

In this paper, we propose an ID based Secure AODV that securely discovers and maintains the route. In our work we have assumed two levels of security: *high* and *low*. By high level of security we mean that, when a path is set up, both the source and the destination node verifies the authenticity of all the other nodes in the route. In addition to this, the authenticity of a node is also verified by its immediate downstream node. In case of *low* level of security, when a path is set up the source and destination node verifies the authenticity of each other (end-to-end) and each intermediate node on the route verifies the authenticity of the downstream node. In addition, we propose an *ID* based secure TCP that securely transmits data using the Diffie-Hellman [9] session key for the MANET nodes. In the proposed scheme, each node has an *ID* which is evaluated from its public key for authentication purpose. Following the proposed scheme a node cannot change its *ID* throughout the lifetime of the MANET. Therefore, the scheme is secure against the above attacks that are associated with AODV and TCP in MANET.

The rest of the paper is organized as follows: Section 2 gives a brief note on the related research efforts in the area of secure routing and TCP based data transmission in MANET. System model followed in this paper is given in Section 3. In Section 4 we present our proposed algorithms namely *ID* based Secure AODV (IDSAODV) and TCP (IDSTCP). Section 5 presents the security analysis and performance of the proposed schemes and finally, conclusions are presented in Section 6.

## 2. RELATED WORK

For providing security in MANET, the main objectives are to make the routing protocol secure and to protect transmitted data. However, these are particularly challenging for MANETs with dynamically changing topologies. Following schemes are proposed in the literature to secure the routing protocol and data transmission of TCP.

Hu et al. [4] have proposed Ariadne, a secure on-demand ad hoc routing protocol based on DSR that prevents attackers or compromised nodes using the symmetric cryptography. To convince the target of the legitimacy of each field in a route request, the initiator simply includes a message authentication code (MAC) in the request. The target can easily verify the authenticity and freshness of the route request using the shared key. One-way hash functions are used to verify that no hop was omitted which is called per-hop hashing. Three alternative techniques to achieve node list authentication: the TESLA protocol [10], digital signatures, and standard MACs.

Secure Routing Protocol (SRP) [5] uses symmetric cryptography to provide end-to-end authentication. The protocol is based on route querying method and it requires a Security Association (SA) between source and destination node. The security association is obtained via the knowledge of the communication counterpart's public key. SRP makes no assumption regarding the intermediate nodes, which exhibits arbitrary and malicious behavior. Nodes use secure message transmission (SMT) [11] to ensure successful delivery of data packets.

The Authenticated Routing for Ad hoc Networks (ARAN) [3] is based on AODV and it is a stand-alone protocol that utilizes cryptographic public-key certificates signed by a trusted authority, which associates its IP address with a public key in order to achieve the security goals of authentication and non-repudiation. ARAN uses cryptographic certificates to bring authentication, message-integrity and non-repudiation to the route discovery process. The source node broadcasts a signed route discovery packet (RDP) to its neighbors for a route to the destination. The RDP includes a packet type identifier, the address of the destination, certificate of the source node, timestamp and a nonce. An intermediate node uses the public key and certificate of its previous node to validate the signature of the RDP. After the validation, it removes the signature of the previous node, appends its own signature and certificate. Similarly, along the reply packet (REP) each node removes signature of its previous node, appends its signature and certificate before forwarding it to the next node. The signature prevents malicious nodes from injecting arbitrary route discovery packets that alter routes or form loops.

Securing AODV (SAODV) [6] proposes a set of extensions that secure the AODV routing packets. Two mechanisms are used to secure the AODV messages: digital signatures to authenticate the non-mutable fields of the messages, and hash chains to secure the hop count information. Since the protocol uses asymmetric cryptography for digital signatures it requires the existence of a key management mechanism that enables a node to acquire and verify the public key of other nodes that participate in the ad hoc network.

The security issues related to transport layer are authentication, securing end-to-end communications through data encryption, handling delays, packet loss and so on. The transport layer protocols in MANET provides end-to-end connection, reliable packet delivery, flow control, congestion control and clearing of end-to-end connection. Though TCP is the main connection oriented reliable protocol in Internet, it does not fit well in MANET. TCP feedback (TCP-F) [12], TCP explicit failure notification (TCPELFN) [12], ad-hoc transmission control protocol (ATCP) [12], and ad hoc transport protocol (ATP) have been developed for MANET. However, none of them have considered the security aspect.

The scheme presented in [13] is based on observation of node mobility. In this scheme, the source node divides the message into multiple shares and sends the shares at different times through different intermediate nodes. The destination node combines the shares to reconstruct the original message. Due to mobility an intermediate node may not be able to collect enough shares to recover the original message. However, it is applicable where delay can be tolerated or the network is dynamic

The SMT scheme is presented in [11] which ensures successful delivery of data packets. In SMT, data messages are divided into multiple packets using secret sharing techniques and sent simultaneously through multiple independent routes. The destination node successfully reconstructs the original message, provided that sufficient shares are received. Each share is transmitted along with message authentication code so that the destination can verify its integrity and the authenticity of its origin. The destination validates the incoming shares and acknowledges the successfully received ones through a cryptographically protected feedback back to the source. However, the scheme assumes that multiple paths exist in the network which may not be true in real scenario.

A popular security mechanism in network layer is IPSEC [14], which is used in wired networks to mitigate most of the attacks discussed in Section 1. IPSEC does not allow an intermediate node to directly access the IP header of a transmitted packet. However, transport layer protocols proposed for ad hoc networks have to rely on information fed back from the intermediate nodes (e.g., *Explicit Congestion Notification* (ECN) [15]), and hence IPSEC cannot be integrated with these protocols [16]. Similar is the case with SSL, PCT and TLS [17] proposed mainly for the wired network.

## 3. SYSTEM MODEL

We consider an autonomous ad hoc network working on its own. It has no gateway or connection to the external world. The network is formed starting from one node and then the other nodes add up one by one similar to IDDIP [18]–[20]. We assume that a node, *A*, have two types of self generated RSA-based key pairs: (1) public (($N_A$; $e_A$))/private ($d_A$) key pair for message

International Journal of Computer Networks & Communications (IJCNC) Vol.6, No.1, January 2014

verification / signing; (2) public ($PK_A$)/ private ($SK_A$) key pair for message encryption/decryption. Here, the node identifier $ID_A$ of node $A$ is generated from its public key (($N_A$; $e_A$)) using a secure one way hash function ($H$). Therefore, a node cannot change its ID within the lifetime of the MANET. In addition, public keys (($N_A$; $e_A$) and $PK_A$) along with identifier $ID_A$ of each node $A$ are distributed before the deployment of the network so that the overhead of the proposed protocol can be reduce. The private keys ($d_A$ and $SK_A$) are kept secret by each node $A$ of the network. Table I presents the notations and their descriptions used in this paper to describe our proposed protocols.

TABLE I

NOTATIONS AND DESCRIPTIONS

| Notations | Description |
|---|---|
| $S$, $D$ and $I$ | source, destination and intermediate nodes |
| $ID_A$ | identifier of node $A$ |
| $IP_A$ | IP address of node $A$ |
| $SN_S$, $SN_D$ | sequence numbers of source and destination |
| $BctID$ | broadcast ID |
| $(N_A, e_A)$ | Public key for signature verification of node $A$ |
| $d_A$ | Private key for signature generation of node $A$ |
| $PK_A$ | Public key for encryption of node $A$ |
| $SK_A$ | Secret key for decryption of node $A$ |
| $K_{AB}$ | Session secret key shared between node $A$ and node $B$ |
| $\sigma_A$ | Digital signature of node $A$ |
| $\delta_A$ | Message Authentication Code tag of node $A$ |

## 4. THE ALGORITHMS

As discussed in Section 2, most of the RSA public key cryptography based secure routing protocols of MANET need to send a large sized public key or certificate along with signature in each routing message. Moreover, these protocols have to rely on *trusted third party* (TTP) for the key and/or certificate distribution to the authorized nodes of the network. In this paper, we develop a RSA-based routing protocol that tries to overcome these problems to a considerable extent by using self-authentication technique. The proposed routing protocol is based on AODV routing protocol. We also observe that the secure routing protocols may not ensure secure data delivery at transport layer of OSI architecture. Here we also present a technique to secure the three-way handshaking process of Transmission Control Protocol (TCP).

*Secure Routing*

We have considered two cases depending upon the level of security: *Case 1:* High and *Case 2:* Low. In Case 1, i,e., for high level of security (*sec_ level* = 1), during routing process, the source and destination nodes separately verify the authenticity of all other nodes in the path. Further, every intermediate node verifies the authenticity of its immediate upstream node. In Case 2, i.e., for low level of security (*sec_level* = 0), during routing process, both the source and the destination nodes verify authenticity of each other. Also each node on the path verifies the authenticity of its immediate upstream node from where it receives the messages (hop-by-hop).

Our proposed routing protocol uses sequential aggregate signatures (SAS) based on RSA [21], [22]. It has two parts: (a) secure route discovery and session key ($K_{AB}$) generation; (b) secure route maintenance.

---

**Algorithm 1:** *IDSAODV Source Node(S)*

1. $d_S \leftarrow$ Get private key for signature generation;
2. $(N_S, e_S) \leftarrow$ Get public key for signature verification;
3. $PK_S \leftarrow$ Get public key for encryption;
4. $SK_S \leftarrow$ Get private key for decryption;
5. $ID_S = H((N_S, e_S))$;
6. Generate $prime\_number\ p$, $random\_number\ r1, g$;
7. $R1 = g^{r1} \pmod{p}$;
8. $R2 = E_{PK_D}(R1)$;
9. $RREQ_S(IP_S, ID_S, SN_S, BctID, R2, p, g, IP_D)$;
10. $h_S = H(RREQ_S, (N_S, e_S))$, $\sigma_S = h_S^{d_S} \pmod{N_S}$;
11. send $(RREQ_S + \sigma_S)$ message to 1-hop neighbours;
12. **if** $sec\_level == 1$ **then**
13.     **if** $(RREP_t + \sigma_t)$ *message is received* **then**
14.         **for** $i \leftarrow t$ to 1 **do**
15.             $ID_G = H((N_i, e_i))$;
16.             $h_i = H(RREP, (N_i, e_i))$;
17.             **if** $\sigma_i^{e_i} \pmod{N_i} = h_i$ AND $ID_G = ID_i$ **then**
18.                 $\acute{\sigma}_{(i-1)} = \sigma_i^{e_i} - h_i \pmod{N_i}$;
19.                 $\sigma_{(i-1)} = \acute{\sigma}_{(i-1)} + b_i N_i$;
20.             **end**
21.         **end**
22.         $ID_G = H((N_D, e_D))$;
23.         $h_D = H(RREP, (N_D, e_D))$;
24.         **if** $\sigma_0^{e_D} \pmod{N_D} = h_D$ AND $ID_G = ID_D$ **then**
25.             Path is secured;
26.             $R = D_{SK_S}(R4)$;
27.             $K_{SD} = R^{r1} \pmod{p}$;
28.         **end**
29.     **end**
30. **end**
31. **else**
32.     **if** $(RREP_t + \sigma_t)$ *message is received* **then**
33.         $ID_t = H((N_t, e_t))$;
34.         $h_t = H(RREP_t, (N_t, e_t))$;
35.         **if** $\sigma_t^{e_t} \pmod{N_t} = h_t$ AND $ID_G = ID_t$ **then**
36.             downstream node $t$ on the route is authenticated;
37.             Extract $\sigma_D$ and $RREP_D$ from $RREP_t$;
38.             $ID_G = H((N_D, e_D))$;
39.             $h_D = H(RREP_D, (N_D, e_D))$;
40.             **if** $\sigma_D^{e_D} \pmod{N_D} = h_D$ AND $ID_G = ID_D$ **then**
41.                 destination node is authenticated;
42.                 $R = D_{SK_S}(R4)$;
43.                 $K_{SD} = R^{r1} \pmod{p}$;
44.             **end**
45.         **end**
46.     **end**
47. **end**

```
Algorithm 2: IDSAODV Intermediate Node(I)
 1  d_I ← Get private key for signature, (N_I, e_I) ← Get public key; ID_I = H((N_I, e_I));
 2  if sec_level == 1 then
 3      if (RREQ_(I-1) + σ_(I-1)) message is received then
 4          ID_G = H((N_(I-1), e_(I-1))), h_G = H(RREQ_(I-1), (N_(I-1), e_(I-1)));
 5          if σ_(I-1)^{e_(I-1)} (mod N_(I-1)) = h_G AND ID_G = ID_(I-1) then
 6              node (I - 1) is authenticated; if σ_(I-1) ≥ N_I then
 7                  σ_(I-1) = σ_(I-1) - N_I; b_I = 1;
 8              end
 9              else
10                  b_I = 0;
11              end
12              update and insert ID_I in RREQ_I message;
13              h_I = H(RREQ_I, (N_I, e_I)), σ_I = (σ_(I-1) + h_I)^{d_I} (mod N_I);
14              send (RREQ_I + σ_I) message to 1-hop neighbours;
15          end
16      end
17      if (RREP_(I-1) + σ_(I-1)) message is received then
18          ID_G = H((N_(I-1), e_(I-1))), h_G = H(RREP_(I-1), (N_(I-1), e_(I-1)));
19          if σ_(I-1)^{e_(I-1)} (mod N_(I-1)) = h_G AND ID_G = ID_(I-1) then
20              node (I - 1) is authenticated; if σ_(I-1) ≥ N_I then
21                  σ_(I-1) = σ_(I-1) - N_I; b_I = 1;
22              end
23              else
24                  b_I = 0;
25              end
26              update and insert ID_I in RREP_I message;
27              h_I = H(RREP_I, (N_I, e_I)), σ_I = (σ_(I-1) + h_I)^{d_I} (mod N_I);
28              send (RREP_I + σ_I) message to upstream node;
29          end
30      end
31  end
32  else
33      if (RREQ_(I-1) + σ_(I-1)) message is received then
34          ID_G = H((N_(I-1), e_(I-1))), h_G = H(RREQ_(I-1), (N_(I-1), e_(I-1)));
35          if σ_(I-1)^{e_(I-1)} (mod N_(I-1)) = h_G AND ID_G = ID_(I-1) then
36              node (I - 1) is authenticated; remove σ_(I-1), ID_(I-1); extract σ_S;
37              insert ID_I, update RREQ_I message;
38              h_I = H(RREQ_I, (N_I, e_I)), σ_I = (σ_S + h_I)^{d_I} (mod N_I);
39              send (RREQ_I + σ_I) message to 1-hop neighbours;
40          end
41      end
42      if (RREP_(I-1) + σ_(I-1)) message is received then
43          ID_G = H((N_(I-1), e_(I-1))), h_G = H(RREP_(I-1), (N_(I-1), e_(I-1)));
44          if σ_(I-1)^{e_(I-1)} (mod N_(I-1)) = h_G AND ID_G = ID_(I-1) then
45              node (I - 1) is authenticated; remove σ_(I-1), ID_(I-1), extract σ_D;
46              insert ID_I, update RREP_I message;
47              h_I = H(RREP_I, (N_I, e_I)), σ_I = (σ_D + h_I)^{d_I} (mod N_I);
48              send (RREP_I + σ_I) message to upstream node;
49          end
50      end
51  end
```

*Secure Route Discovery and Session Key Generation: Case 1*: When we need high level of security i.e., *sec_level* = 1, the secure route discovery procedure of the proposed protocol works as follows: to create a path between a source node *S* and a destination node *D*, the source node, *S*, first generates a prime number *p* along with two random numbers *r*1 and *g*, where *p* and *g* are publicly known parameters. *S* then computes $R1 = g^{r1}$ (*mod p*), encrypts $R2 = E_{PKD}(R1)$ broadcasts it in a signed ($σ_S$) $RREQ_S$ message along with $ID_S$ to its neighbours. The $RREQ_S$ message also contains source IP $IP_S$, source sequence number *SNS*, broadcast ID *BctID*, and destination IP $IP_D$ as similar to AODV protocol.

An intermediate node, *I*, on receiving the signed ($σ_{(I-1)}$)$RREQ_{(I-1)}$ message from node (*I-1*) first checks the authenticity of the node (*I - 1*). If node (*I - 1*) is authenticated, node *I* inserts its ID $ID_I$ and subsequently updates the $RREQ_I$. The intermediate node *I* also generates an aggregate signature ($σ_I$) from both $RREQ_I$ message and the received signature ($σ_{(I-1)}$). Thereafter node *I* broadcasts $RREQ_I$ message along with the aggregate signature $σ_I$ to its neighbours. This process continues till the *RREQ* message is received by the destination node.

```
Algorithm 3: IDSAODV Destination Node(D)
 1  d_D ← Get private key for signature generation; (N_D, e_D) ← Get public key for signature verification;
 2  PK_D ← Get public key for encryption; SK_D ← Get private key for decryption;
 3  ID_D = H((N_D, e_D));
 4  if sec_level == 1 then
 5      if (RREQ_t + σ_t) message is received then
 6          for i ← t to 1 do
 7              ID_G = H((N_i, e_i));
 8              h_i = H(RREQ_i, (N_i, e_i));
 9              if σ_i^{e_i} (mod N_i) = h_i AND ID_G = ID_i then
10                  σ̂_(i−1) = σ_i^{e_i} − h_i (mod N_i);
11                  σ_(i−1) = σ̂_(i−1) + b_i N_i;
12              end
13          end
14          ID_G = H((N_S, e_S));
15          h_S = H(RREQ_S, (N_S, e_S));
16          if σ_0^{e_S} (mod N_S) = h_S AND ID_G = ID_S then
17              Path is secured;
18              Generate random number r2; R3 = g^{r2} (mod p);
19              R4 = E_{PK_S}(R3);
20              Generate RREP_D(IP_S, ID_S, BctID, SN_D, IP_D, ID_D, R4) message;
21              h_D = H(RREP_D, (N_D, e_D)); σ_D = h_D^{d_D} (mod N_D);
22              send (RREP_D + σ_D) message to upstream node;
23              R = D_{SK_D}(R2); K_{DS} = R^{r2} (mod p);
24          end
25      end
26  end
27  else
28      if (RREQ_t + σ_t) message is received then
29          ID_t = H((N_t, e_t));
30          h_t = H(RREQ_t, (N_t, e_t));
31          if σ_t^{e_t} (mod N_t) = h_t AND ID_G = ID_t then
32              upstream node (previous hop) t on the route is authenticated;
33              extract σ_S and RREQ_S from RREQ_t;
34              ID_S = H((N_S, e_S));
35              h_S = H(RREQ_S, (N_S, e_S));
36              if σ_S^{e_S} (mod N_S) = h_S AND ID_G = ID_S then
37                  source node is authenticated;
38                  Generate random number r2;
39                  R3 = g^{r2} (mod p);
40                  R4 = E_{PK_S}(R3);
41                  RREP_D(IP_S, ID_S, BctID, SN_D, IP_D, ID_D, R4);
42                  h_D = H(RREP_D, (N_D, e_D));
43                  σ_D = h_D^{d_D} (mod N_D);
44                  send (RREP_D + σ_D) message to upstream node;
45                  R = D_{SK_S}(R4);
46                  K_{DS} = R^{r1} (mod p);
47              end
48          end
49      end
50  end
```

On receiving the signed ($\sigma_t$) $RREQ_t$ message, the destination node $D$ first checks the authenticity of all the intermediate nodes $ID_t$ including source node $ID_S$ on the route. It also checks the authentication of the received aggregate signature $\sigma_t$ by verifying all the signatures of node $S$ to node $t$. If both checks pass, the destination $D$ decrypts $R = D_{SKD}(R2)$ and generates the session key $K_{DS} = R^{r2}$ (*mod p*). It also generates a random number $r2$ and computes $R3 = g^{r2}$ (*mod p*). $D$ thereafter encrypts $R4 = E_{PKS}(R3)$ and unicasts it in a signed ($\sigma_D$) $RREP_D$ message with its $ID_D$ to $S$ along the reverse direction of $RREQ$ message. The $RREP_D$ message also contains other parameters of AODV (such as, source IP $IP_S$, destination sequence number $SN_D$). An intermediate node verifies the authenticity of the $RREP$ message and combines its signature with the signatures of previous hops on the route in the same way as $RREQ$ message. When $S$ receives the signed ($\sigma_t$) $RREP_t$ message, it checks the authenticity of each node including $D$ on the route by verifying all the *ID*s and $\sigma$s. If both checks pass, $S$ decrypts $R = D_{SKS}(R4)$ and generates the session key $K_{SD} = R^{r1}$ (*mod p*) to send the data packets to $D$ via this route. Figure 2 shows an example of the route discovery process of our proposed routing protocol for *Case 1*.

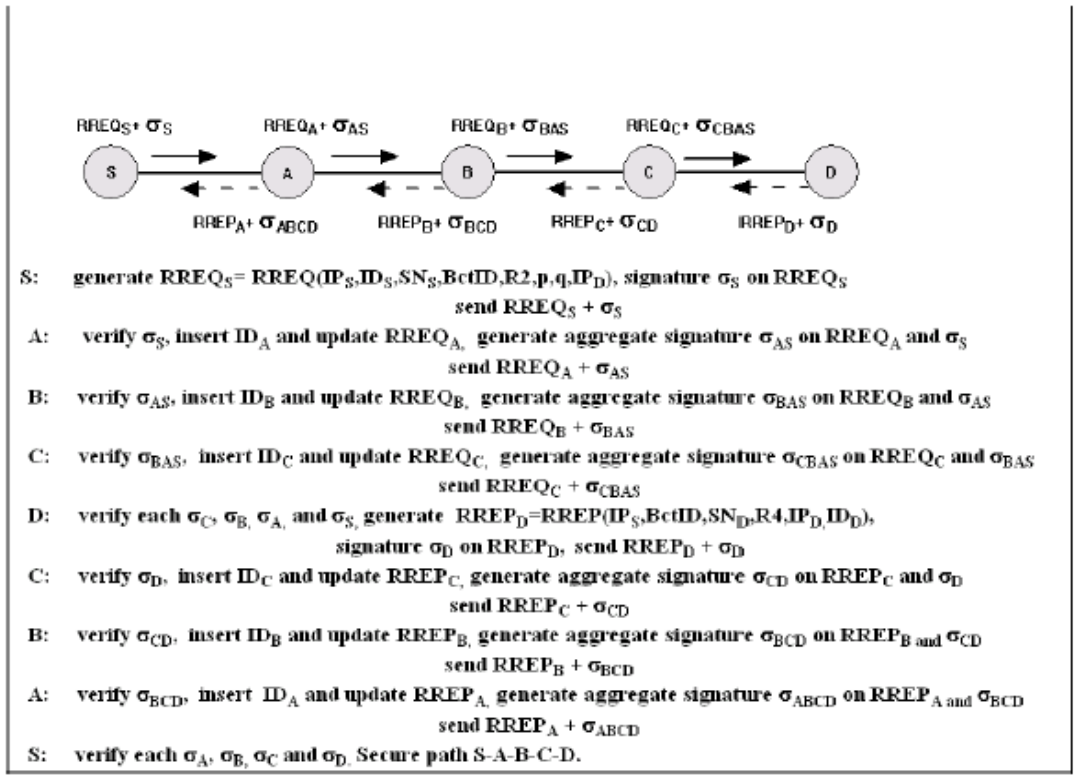

Fig. 2. An example of secure route discovery process for *Case 1*

*Case 2*: For low security level, i.e. *sec_level* = 0, initially the source node broadcasts signed $\sigma_S$ $RREQ_S$ message along with its ID, $ID_S$, to the neighbours in a way similar to the previous case. An intermediate node, *I*, on receiving the signed $\sigma_{(I-1)}$ $RREQ_{(I-1)}$ message from node (*I-1*), first checks the authenticity of the node (*I-1*). If the node (*I-1*) is authenticated, it removes the signature of the node (*I-1*), inserts $ID_I$ and updates the $RREQ_I$ message. It also generates its own signature on the $RREQ_I$ message and the signature $\sigma_S$ of *S* and broadcasts it to its neighbours. This process continues till the *RREQ* message reaches the destination node. On receiving the signed $\sigma_t$ $RREQ_t$ message, the destination node *D* first checks the authenticity of the node $ID_t$ and the source node $ID_S$. It also checks the authentication of the received signature $\sigma_t$ (signature of the node *t* from whom it receives *RREQ*) and $\sigma_S$ of the source node. If both checks pass, *D* generates the session key $K_{DS}$ and unicasts the signed ($\sigma_D$) $RREP_D$ message with its $ID_D$ to *S* along the reverse direction of *RREQ* message in the same way as in the first case. An intermediate node *I* verifies the signature $\sigma_{(I-1)}$ of the received $RREP_{(I-1)}$ message. If checks pass, it removes the signature $\sigma_{(I-1)}$ and ID $ID_{(I-1)}$, and inserts its own ID $ID_I$ and subsequently updates $RREP_I$ message. It also generates signature $\sigma_I$ on $RREP_I$ and the signatures $\sigma_D$ of *D*. When *S* receives the signed ($\sigma_t$) *RREP* message, it checks the authenticity of the previous node *t* and *D* on the route by verifying the *ID* and $\sigma$ of both the nodes. If both checks pass, *S* generates the session key $K_{SD}$ to send the data packets to *D* via this route. An example of the route discovery process of our proposed routing protocol for *Case 2* is shown in Figure 3. The route discovery process for source, intermediate and destination nodes are presented in Algorithm 1, Algorithm 2 and Algorithm 3 respectively.

*Secure Route Maintenance: Case 1:* For high level of security, i.e., *sec_level* = 1, the proposed protocol maintains a established route as follows: If a node *X* detects that its immediate down link towards *D* is broken, it sends signed ($\sigma_X$) $RERR_X$ message with $ID_X$ along the reverse route toward

*S*. On receiving a signed ($\sigma_{(I-1)}$) *RERR$_{(I-1)}$* message from node (*I-1*), an intermediate node *I* immediately checks the authenticity of the node (*I-1*) by verifying the signature $\sigma_{(I-1)}$ and *ID$_{(I-1)}$*.

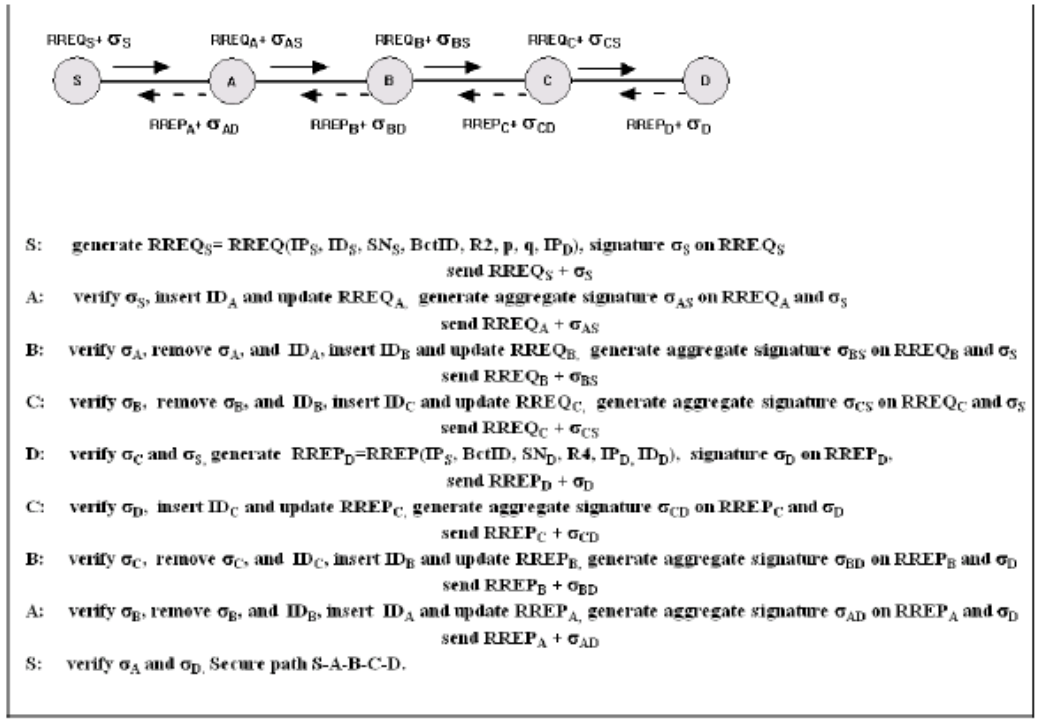

Fig. 3. An example of secure route discovery process for *Case 2*

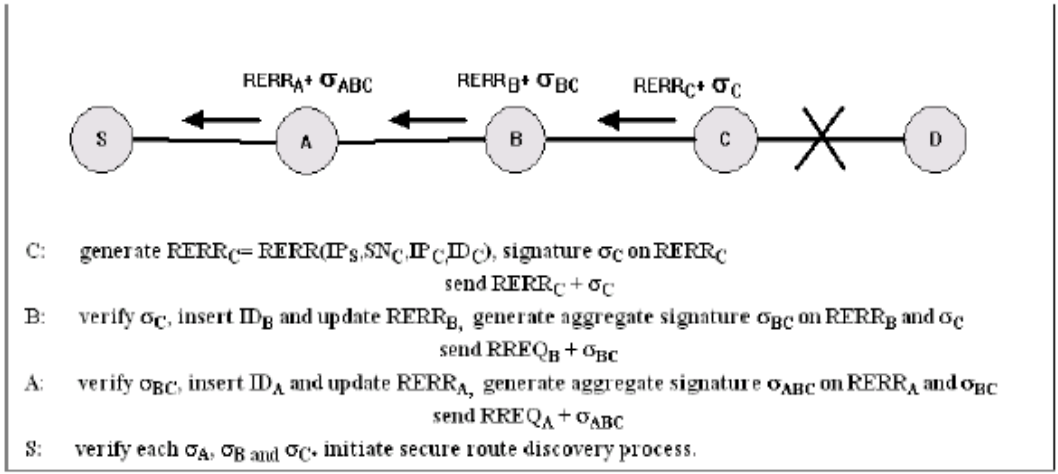

Fig. 4. An example of secure route maintenance process for *Case 1*

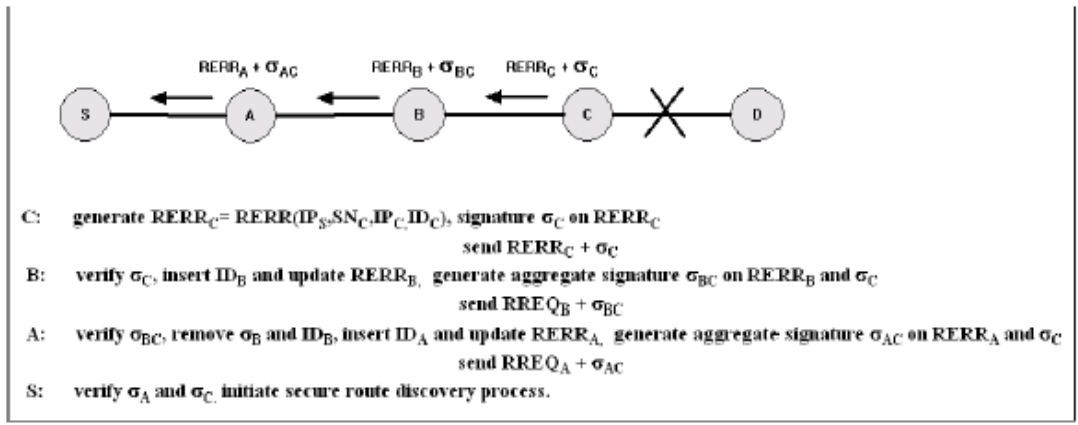

Fig. 5. An example of secure route maintenance process for *Case 2*

If node ($I$-$1$) is authenticated, it inserts $ID_I$ and updates the $RERR_I$ message. The node $I$ also generates a signature $\sigma_I$ from the $RERR_I$ message and the received signature $\sigma_{(I-1)}$ and forwards the signed $\sigma_I$ $RERR_I$ message along the path toward $S$. On receiving the $\sigma_t$ $RERR_t$ from $t$, $S$ verifies all the signatures and $ID$s of the nodes on the route. If both verifications pass, $S$ initiates the route discovery process of our proposed routing protocol. The secure route maintenance process for *Case 1* of our proposed routing protocol is given by an example in Figure 4.

*Case 2:* For low level of security, i.e., *sec_level* = 0, the route maintenance of the proposed protocol works as follows: After detecting the connection loss, node $X$ sends signed ($\sigma_X$) $RERR_X$ message with $ID_X$ along the reverse path toward $S$. On receiving the signed ($\sigma_{(I-1)}$) $RERR_{(I-1)}$ message from node ($I - 1$), an intermediate node $I$ first checks the authenticity of the node ($I - 1$) by verifying the signature $\sigma_{(I-1)}$ and $ID_{(I-1)}$. If the node ($I - 1$) is authenticated, it removes $ID_{(I-1)}$ and updates the $RERR_I$ message by appending $ID_I$. It also generates the signature $\sigma_I$ on the $RERR_I$ message and the signature $\sigma_X$ of node $X$. Node $I$ forwards the signed $\sigma_I$ $RERR_I$ message along the path toward $S$. On receiving the $\sigma_t$ $RERR_t$ from $t$, $S$ verifies the signatures $\sigma_t$, $\sigma_X$ and $ID_t$, $ID_X$. If both verifications pass, $S$ initiates the route discovery process. The secure route maintenance process for *Case 2* of our proposed routing protocol is shown by an example in Figure 5. The algorithm for secure route maintenance process of our proposed routing protocol is given in Algorithm 4.

*Secure Data Transmission*

As discussed in the previous Section 4, after discovering the secure path, source and destination node have common session secret key (i.e., $K_{SD}$= $K_{DS}$). Initially source node ($S$) starts connection establishment with destination node ($D$) using three-way handshaking of TCP. $S$ at first generates the initial sequence number ($ISN_S$) from a random number ($R$) and a hash function of source port, destination port, $ID_S$, $ID_D$ and session secret key $K_{SD}$. Subsequently, it generates authentication tag ($\delta_S$) on $SYN(ISN_S)$ segment using HMAC function and $K_{SD}$, sends it to $D$ along with $SYN(ISN_S)$ segment. On receiving the $SYN(ISN_S)$ + $\delta_S$, $D$ generates the authentication tag ($\delta_G$) from the received $SYN(ISN_S)$ and $K_{DS}$. If the generated tag ($\delta_G$) and received tag ($\delta_S$) are same, $S$ is authenticated to $D$. At this point $D$ also generates the initial sequence number ($ISN_D$) and authentication tag ($\delta_D$) on $SYN(ISN_D)$ + $ACK(ISN_S+1)$ segment in a way similar to $S$, and sends it to $S$ along with the segment. On receiving $SYN(ISN_D)$ + $ACK(ISN_S+1)$ + $\delta_D$, $S$ generates the authentication tag ($\delta_G$) on $SYN(ISN_D)$ +$ACK(ISN_S+1)$ segment and matches the generated tag ($\delta_G$) with the received tag ($\delta_D$). If both are same, $D$ is authenticated to $S$, and $ACK(ISN_D+1)$ + $\delta_S$ segment is sent by $S$. $D$ generates the authentication tag ($\delta_G$) on the received $ACK(ISN_D+1)$ segment and checks it with the received tag ($\delta_S$). If the tags match, $S$ is authenticated. This completes the three-way handshake process and therefore $D$ allocates the resource for $S$ to start transmission of data along with the authentication tag.

The algorithms for the three-way handshake connection establishment process for source ($S$) and destination ($D$) nodes are given in Algorithm 5 and Algorithm 6 respectively. Figure 6 shows a schematic for a secure three-way handshaking connection establishment process of our proposed protocol using a timing diagram.

```
Algorithm 4: IDSAODV Route Maintenance
 1 if node X detects link break then
 2     RERR_X; h_X = H(RERR_X, (N_X, e_X)), σ_X = h_X^{d_X} (mod N_X);
 3     send (RERR_X + σ_X) message to upstream node;
 4 end
 5 if sec_level = 1 then
 6     if upstream node I receives (RERR_{(I-1)} + σ_{(I-1)}) message then
 7         ID_G = H((N_{(I-1)}, e_{(I-1)})); h_{(I-1)} = H(RERR_{(I-1)}, (N_{(I-1)}, e_{(I-1)}));
 8         if σ_{(I-1)}^{e_{(I-1)}} (mod N_{(I-1)}) = h_{(I-1)} AND ID_G = ID_{(I-1)} then
 9             node (I − 1) is authenticated; if σ_{(I-1)} ≥ N_I then
10                 σ_{(I-1)} = σ_{(I-1)} − N_I; b_I = 1;
11             end
12             else
13                 b_I = 0;
14             end
15             insert ID_I in RERR_I; h_I = H(RERR_I, (N_I, e_I)); σ_I = (σ_{(I-1)} + h_I)^{d_I} (mod N_I);
16             send (RREQ_I + σ_I) message to upstream node;
17         end
18     end
19     if node S receives (RERR_t + σ_t) message then
20         for i ← t to 1 do
21             ID_G = H((N_i, e_i)); h_i = H(RERR_t, (N_i, e_i));
22             if σ_i^{e_i} (mod N_i) = h_i and ID_G = ID_i then
23                 σ́_{(i−1)} = σ_i^{e_i} − h_i (mod N_i); σ_{(i−1)} = σ́_{(i−1)} + b_i N_i;
24             end
25         end
26         ID_G = H((N_X, e_X)); h_X = H(RERR_X, (N_X, e_X));
27         if σ_0^{e_X} (mod N_X) = h_X AND ID_G = ID_X then
28             X is authenticated;
29         end
30     end
31 end
32 else
33     if upstream node I receives (RERR_{(I-1)} + σ_{(I-1)}) message then
34         ID_G = H((N_{(I-1)}, e_{(I-1)})); h_{(I-1)} = H(RERR_{(I-1)}, (N_{(I-1)}, e_{(I-1)}));
35         if σ_{(I-1)}^{e_{(I-1)}} (mod N_{(I-1)}) = h_{(I-1)} AND ID_G = ID_{(I-1)} then
36             (I − 1) authorized; remove σ_{(I-1)}, ID_{(I-1)}; extract σ_X; insert ID_I, update RERR_I;
37             h_I = H(RERR_I, (N_I, e_I)), σ_I = (σ_X + h_I)^{d_I} (mod N_I);
38             send (RREQ_I + σ_I) message to upstream node;
39         end
40     end
41     if node S receives (RERR_t + σ_t) message then
42         ID_G = H((N_t, e_t)); h_t = H(RERR_t, (N_t, e_t));
43         if σ_t^{e_t} (mod N_t) = h_t AND ID_G = ID_t then
44             downstream node t is authenticated; extract σ_X and RERR_X from RERR_t;
45             ID_X = H((N_X, e_X)); h_X = H(RERR_X, (N_S, e_S));
46             if σ_X^{e_X} (mod N_X) = h_X AND ID_G = ID_X then
47                 node X is authenticated;
48             end
49         end
50     end
51 end
```

The above process is followed to secure the three-way handshake connection termination process too. Session key $K_{SD}$ terminates after the end of one session or at any stage if authentication fails in the three-way handshake process. For a new session, a new key is obtained at the time of route discovery and the process is repeated.

## 5. SECURITY AND PERFORMANCE ANALYSIS

In our proposed protocol the identifier *ID* of a node is generated from its public key using a secure one-way hash function. In addition, the public keys and *ID* of the nodes are distributed before the deployment of the MANET. As a result, a node cannot change its *ID* within the lifetime of a MANET. Therefore, impersonation attacks or unauthorized participation is not possible. In our proposed secure routing protocol, as the source/destination node signs the *RREQ*/ *RREP* packet using the private key of the source/ destination node, a malicious node will not be able to modify the route sequence number. As *RREQ*/ *RREP* packets are signed by the each intermediate node, a malicious node cannot change the value of the hop count field. Due to similar reason, a malicious node cannot change the source/ destination address of a packet. In the first scenario, node *ID*s are cached and the routing messages are signed by them. The source/ destination node verifies the authenticity of each node on the route. In the second scenario, authentication is done in hop-by-hop and end-to-end manner. Therefore, two or more malicious nodes cannot collaborate to make the tunnel. In the proposed scheme, messages are signed by the nodes on the route at the time of route maintenance. Therefore, a malicious node cannot falsely inject route error message to the source.

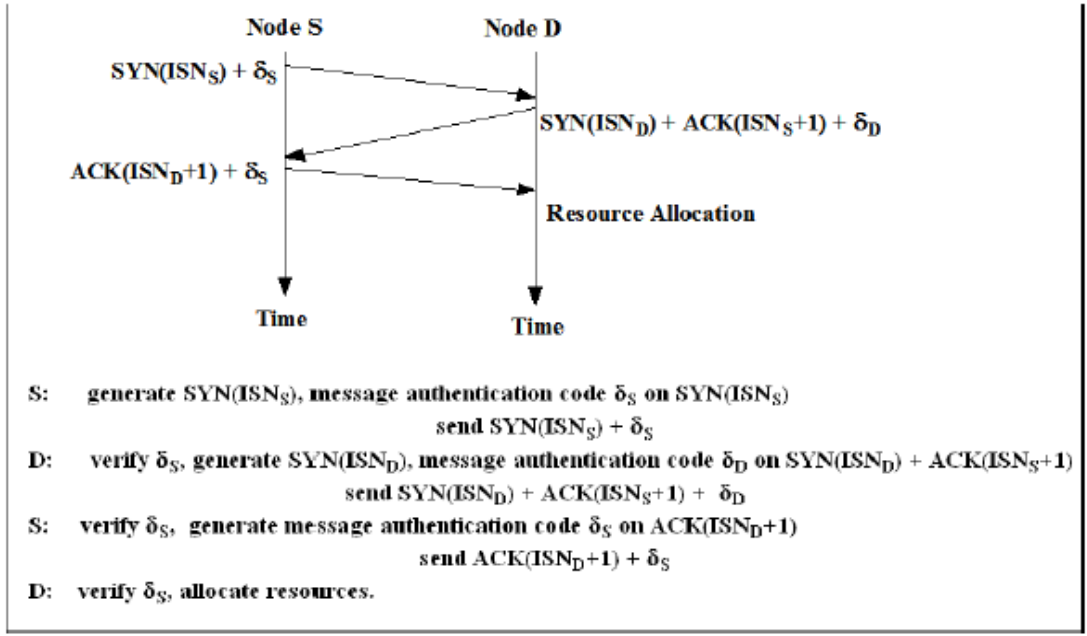

Fig. 6. Timing diagram of secure three-way handshaking connection establishment process

In our proposed secure transmission control protocol, all the segments are sent along the *message authentication code* (MAC) [23] tags. A MAC tag is a short piece of information used to authenticate a message and to provide integrity and authenticity assurances on the message. Integrity assurances detect accidental and intentional message changes, while authenticity assurances affirm origin of the message. Here, the MAC tag is generated using a hash function, a session secret key and an arbitrary-length message. Further, a node cannot hide and change its *ID*

```
Algorithm 6: IDSTCP Destination Node
1  if (SYN(ISN_S) + δ_S) message is received from source node then
2      δ_G = H(SYN(ISN_S), K_DS); //Session key K_DS
3      if δ_G == δ_S then
4          ISN_D = R + H(src_port, dst_port, ID_S, ID_D, K_DS);
5          SYN(ISN_D) + ACK(ISN_S + 1);
6          δ_D = H(SYN(ISN_D) + ACK(ISN_S + 1), K_DS);
7          send (SYN(ISN_D) + ACK(ISN_S + 1) + δ_D) message to source node;
8      end
9      else
10         Drop SYN(ISN_S) message;
11     end
12 end
13 if (ACK(ISN_D + 1) + δ_S) message is received from source node then
14     δ_G = H(ACK(ISN_D + 1), K_DS);
15     if δ_G == δ_S then
16         Connection established;
17         Allocate resources;
18     end
19     else
20         Drop ACK(ISN_D + 1) message;
21     end
22 end
```

or the *ID* cannot be spoofed by any other nodes in the network. This makes the *SYN* flooding attack difficult for an attacker. Initial sequence number (*ISN*) is generated by the source and the destination nodes from a monotonically increasing random number (*R*) and a hash function of source port, destination port, node identifiers of the source and destination nodes, and a session secret key $K_{SD}$. Therefore, a malicious node may not be able to guess the initial sequence number (*ISN*) and therefore cannot hijack the session or launch *ACK* storm in the MANET.

As the scheme uses either signature (in case of secure routing protocol) or *MAC* (in case of secure transmission control protocol) for authentication, message forging is not possible. Therefore, the attacks associated with AODV and TCP are not possible. An attack may be possible if an attacker is somehow able to generate a public/ private key pair that is exactly similar to an authorized node in the network. This probability is $1=2^k$, where *k* is the size of the public key. For *k* = 1024 bits, it can be seen that the probability is extremely low and is almost impossible within a small life time of MANET.

As our proposed protocol uses identity-based cryptography, it does not need certificate for authentication. Further, public keys along with *ID*s of the nodes are distributed before the deployment of the MANET. Thus it eliminates the need of sending the public key with the packet. Since the SAS-based scheme is based on plain RSA, its per-signer signature generation cost is equivalent to that of a plain RSA signature. However, following are the overheads of our proposed protocols:

- Each node has to maintain the list of public keys along with *ID*s of other nodes in the network.
- In *Case 1*, *ID*s and signatures of all the nodes in the route are required to be cached. However, the signature expands by one bit per signers that is if the number of signers is *t*, then the signature expands by *t*-bit here.
- Each node has to verify its predecessor and sign the messages. Source/destination node has to verify all the nodes on the route in *Case 1*.
- To generate the session key for a session, each time encryption and decryption has to be done by the source node and the destination node respectively.

It may be noted here that our proposed protocol may not detect attacks from internal or compromised malicious nodes. It requires trust based protocol [24] as a second wall of defence to detect attacks by internal malicious nodes.

## 6. CONCLUSION

In this paper the popular MANET routing protocol AODV and the standard TCP has been improved and made suitable for using it in mobile ad hoc networks. The proposed routing protocol provides security to the route discovery and route maintenance phases. Further, the three-way handshaking process of standard TCP has been secured. Here each node is made to have an *ID* that is generated from its public key and is unchangeable throughout the lifetime of the network. Performance analysis shows that our proposed protocols are secure against the attacks that are associated with AODV and TCP in MANET.

**AUTHORS**

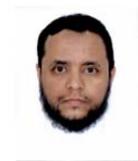

Waleed S. Alnumay received his bachelor degree in Computer Science from King Saud University, Riyadh, Saudi Arabia in the year 1993. He did his master degree in Computer Science from University of Atlanta, Atlanta, Georgia, USA in the year 1996. He completed his PhD in Computer Science from Oklahoma University, Norman, Oklahoma, USA in the year 2004. Dr. Alnumay is currently working as an Assistant Professor of Mobile Networking and vice dean of eLearning deanship in Computer Science Department, King Saud University. He has published research papers in reputed international conferences and journals. His main research interest is Computer Networks and Distributed Computing that includes but not limited to Mobile Ad-hoc and Sensor Networks, Information-Centric Networking and Software-Defined Networking.

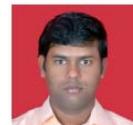

Uttam Ghosh received his B. Tech. in Information Technology from Govt. College of Engineering and Textile Technology, Serampore, WB, India in 2005. He completed his MS (by Research) and Ph.D both from the Department of Electronics and Electrical Engineering, Indian Institute of Technology Kharagpur, India in 2009 and 2013 respectively. Dr. Ghosh has a number of publications in International Journals (including Elsevier Ad Hoc Networks, Wiley NEM, IETE Journal of Research, IET Networks) and Conferences (including IEEE N2S, WCNC, IPDPS and ICDCN). His main research interests include Computer Networks, Security and Cryptography, Distributed and Trust Computing, Wireless Networks, Mobile Ad-hoc Networks, Information-Centric Networking and Software-Defined Networking.